# Recent Sharp Rise in Inhomogeneous Hydrological Extremes Stress Vegetation Growth in China

Shengyuan Liu[1,2,3†], Jeremy Cheuk-Hin Leung[1,4†], Jianjun Xu[2,3,5*], Kunlun Xiang[6], Shifei Tu[2,3,5], Meiying Zheng[7], Daosheng Xu[1], Yi Li[1], Banglin Zhang[1,4,8,9*]

Affiliations

[1]College of Meteorology and Oceanography, National University of Defense Technology, Changsha, China

[2]College of Ocean and Meteorology, Guangdong Ocean University, Zhanjiang, China

[3]Western Guangdong Key Laboratory of Marine Meteorological Disaster Theory and Application, Guangdong Ocean University, Zhanjiang, China

[4]Laboratory of Atmospheric Environmental Monitoring and Early Warning for Low-Altitude Economy, National University of Defense Technology, Changsha, China

[5]Shenzhen Institute of Guangdong Ocean University, Shenzhen, China

[6]Guangdong Ecological Meteorological Centre, Guangzhou, China

[7]Department of Atmospheric Sciences, Yunnan University, Kunming, China

[8]College of Atmospheric Sciences, Lanzhou University, Lanzhou, China

[9]Key Laboratory of High Impact Weather (Special), China Meteorological Administration, Changsha, China

* Correspondence: Banglin Zhang (zhangbanglin24@nudt.edu.cn), Jianjun Xu (jxu@gdou.edu.cn)

† These authors contributed equally to this work.

**Abstract**

The intensifying spatial inhomogeneity of rainfall under greenhouse warming implies that more inhomogeneous hydrological extremes (IHEs), i.e., coexistence of extreme rainfall or drought, may be triggered. While vegetation growth in China is sensitive to hydrological hazards, the variability of IHEs and their ecological impacts remain underexplored. Here, we find a significant increase in IHEs during China's growing season since 2000 (+2.1 events or +14.52 days per decade), with a rapid sharp rise to an annual average of 6.4 events or 42.0 days in the past five years. The primary driver is the enhanced inhomogeneity of moisture-dynamic coupled weather conditions, overlapping with a northward shift of climatological precipitation distribution. This results in a "Wet-North and Dry-South" pattern of IHE impacts, which poses severe and asymmetric threats to vegetation growth in China, with the expansion of drought areas exerts stronger stress on vegetation than the compensatory effects of rainfall. Our findings suggest that the sharp rises of IHEs tend to yield net negative impacts on vegetation growth, highlighting the need for stronger hydrological management to reduce future risks.

## Introduction

The warming climate has driven significant increases in the frequency and intensity of extreme precipitation events, yet these changes have shown pronounced inhomogeneity across regions[1–4]. Precipitation inhomogeneity represents a key dimension beyond mean precipitation, capturing the temporal variability or spatial dispersion features of rainfall. Recent research have documented that the warming climate has increased precipitation variability and the likelihood of extremes, reshaping the probability distribution of precipitation[4,5]. Meanwhile, global precipitation has become more spatially heterogeneous[6], with convective-scale rainfall becoming more concentrated[2]. In addition, the spatial dependence of extremes across regions has grown stronger[7]. These indicate that climate change has led to increasingly inhomogeneous and extreme precipitation patterns, and related climate risks are likely to escalate in the future.

A potential sufferer of such enhanced spatial inhomogeneity in precipitation is vegetation ecology. Although the $CO_2$ fertilization effect has generally enhanced vegetation productivity in the past decades[8–10], the coexistence of global greening with localized browning highlights the increasing prevalence of climatic stressors[9,11,12], such as drought[13–15]. These events constrain growth or induce mortality, leaving legacy effects that compromise ecosystem resilience to subsequent perturbations[16–18]. Crucially, shifts in precipitation heterogeneity, rather than average rainfall, are more closely related to hydrological extremes[19,20]. When such extremes occur synchronously across large regions, they often accumulate over time or space, overwhelming the capacity for seed dispersal and climate refugia to aid recovery, locking ecosystems into degraded states[13,21].

As the inhomogeneity of precipitation intensified, the probability of wet-dry contrasts also increases[2,4,6,19]. Such changes may favor the occurrence of inhomogeneous hydrological extremes (IHEs). Studies discovered that anthropogenic climate forcing has led to more concurrent extremes [7,22,23]. Particularly, this kind of IHE events have impacted China lately. For example, the record-breaking Meiyu brought extreme flooding to the Yangtze River basin in 2020[24,25], and simultaneously led to strong dry anomalies over Southern and Southwestern China[26]. In 2022, on the contrary, the Yangtze River basin suffered from severe droughts[27,28], while Northeastern and Southern China experienced extreme wet conditions[29]. Despite these record-breaking events as well as their impacts, it remains unexplored whether these extreme events have become more frequent in China recently.

IHE events can cause severe declines in vegetation productivity[27], and their risk is of particular concern in China, one of the hotspots contributing to the global greening trend[30,31]. As a major food-producing country, China leads global grain production and maintains an extensive area of arable land[32]. Eastern plains and monsoon zones sustain intensive agriculture, where crop productivity is closely tied to hydrological conditions[33]. In contrast, the arid and semi-arid regions of Northwestern China, as well as the Tibetan Plateau, are characterized by high spatial heterogeneity in both hydrology and landscape[34]. The pronounced spatial disparities in extreme environment across China disproportionately impact regional ecosystems. These contrasts make China more vulnerable to IHEs.

In short, it is hypothesized that the increased precipitation inhomogeneity under climate change may trigger more IHE events, which may directly impact on vegetation, in China. However, to date, quantitative research on IHEs in China, as well as their impacts on vegetation, remains limited. While the impacts of global warming or individual events on vegetation have been extensively studied, a critical gap remains in our understanding of the effects of spatially inhomogeneous extreme event patterns. Accordingly, the key question then becomes: How does the spatial inhomogeneity of hydrological extremes shape vegetation response to recent climate change? To address this, in this study, we investigate recent changes in these IHEs and examine their ecological impacts, with focus put on the growing season (March–November) in China owing to its wide range of climatic conditions, complex topography, and diverse vegetation types. Our results demonstrate a nonlinear rise of IHEs in China since 2000, characterized by increase in occurrence, duration and coverage. Consequently, the vegetation has been subjected to a growing risk of such extreme events, with a notable exacerbation observed over the past five years.

## Results

### Enhancing inhomogeneous precipitation distribution in China

Spatial variance (SV) of 10-day standardized precipitation index (SPI-10d) ($SV_{SPI}$, see Methods) is an effective measure quantifying the degree of spatial inhomogeneity in rainfall[6]. As shown in Fig. 1a, $SV_{SPI}$ exhibits apparent seasonal cycles and interannual fluctuations, with peaks corresponding to episodes of extreme spatial inhomogeneity. Over the past two decades, significant increase in the inhomogeneity of synoptic-scale precipitation has been observed over China, with a significant $SV_{SPI}$ upward tendency of 15.3%/decade during 2000–2024 ($p<0.01$), suggesting that the spatial pattern of wet-dry anomalies has been shifting to a more inhomogeneous state.

Alongside the significant increasing tendency, $SV_{SPI}$ has peaked frequently in recent years, notably in June 2020 and August 2022. Both peak periods are marked by widespread wet anomalies and dry anomalies over China, corresponding to extremely inhomogeneous precipitation distribution. Specifically, the peak in June 2020 corresponds to the famous "violent Meiyu" event over the Yangtze River Basin[24,25], dominated by positive rainfall anomalies over central China and negative anomalies over southern China (Fig. 1b). In contrast, the peak in 2022 is characterized by a large coverage of severe drought over the Yangtze River Basin and anomalously wet conditions over northern China[27,35] (Fig. 1d). Despite their contrasting hydrological pattern, both events coincided with the growing season and the major agricultural regions in China, which poses a potential widespread hazard to ecosystems from such extremes. By locating these $SV_{SPI}$ peaks, one can identify days with extremely inhomogeneous precipitation distribution and define them as IHEs (see Methods).

IHE events have direct and persisting impacts on vegetation, which can be quantified by changes in solar-induced chlorophyll fluorescence (ΔSIF) between the pre-event and post-event periods (see Methods). Taking the two IHE events in June 2020 and August 2022 as examples, the small initial positive vegetation anomaly during both events is rapidly followed by a persistent negative response relative to the baseline. The vegetation stress continued after the events ended, delaying recovery of vegetation for several months (Figs. 1c and 1e). These two case studies both reveal that IHEs can trigger potential threats to vegetation, that may last for several months or even cover the whole growing seasons. If such IHE events become more frequent, their cumulative impact would extend far beyond the localized damage caused by individual rainfalls or droughts. The observed increase in $SV_{SPI}$ over China makes this a realistic concern. To verify this possibility, we first answer the question: are IHEs becoming more common in recent years?

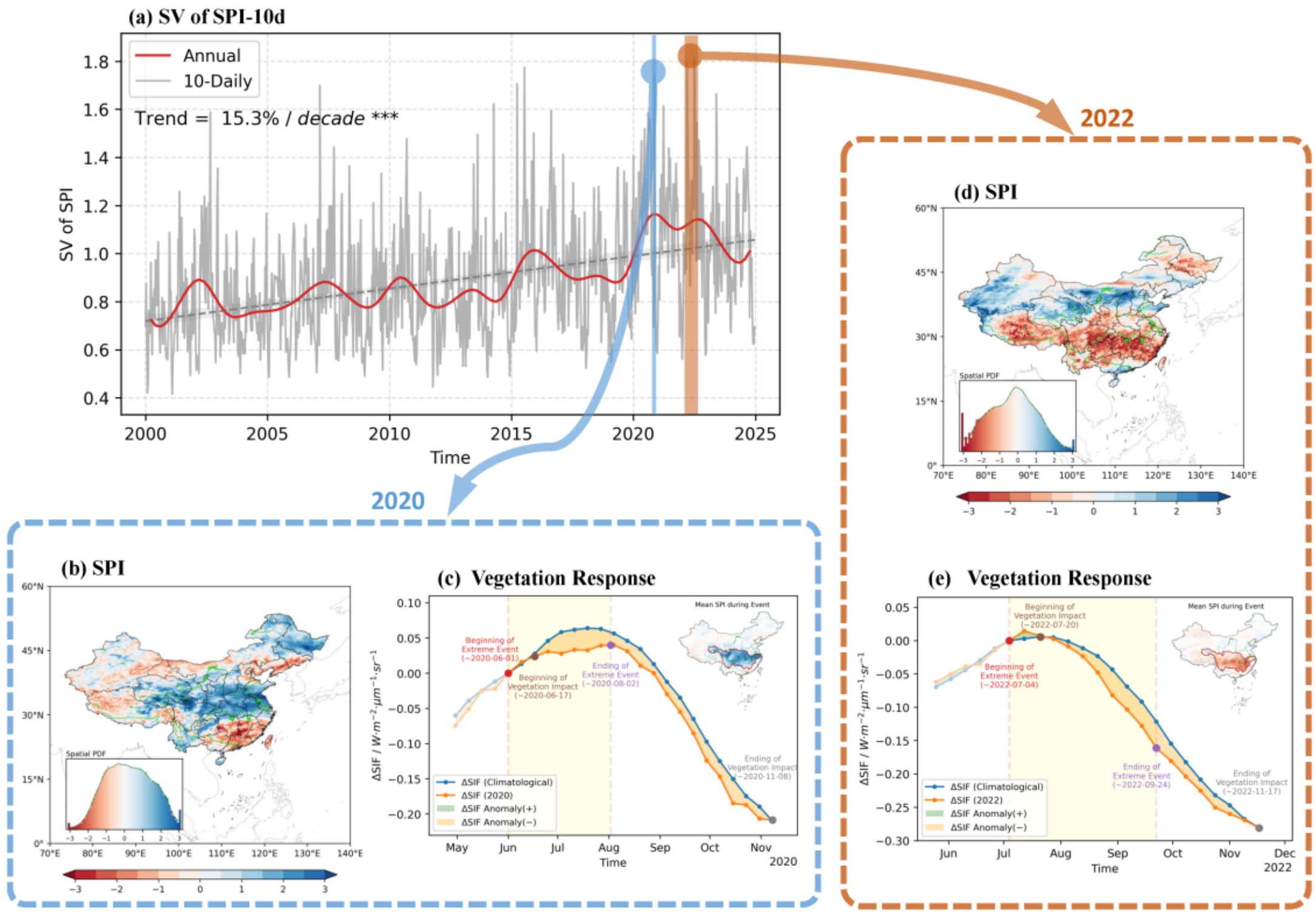

**Fig.1 | Inhomogeneous hydrological extreme events. (a)** Time series of $SV_{SPI}$ (units: dimensionless) in China from 2000 to 2024, with the gray line showing the $SV_{SPI}$ values at 10-day intervals, and the red line showing the annual mean $SV_{SPI}$. **(b)** Spatial distribution of SPI-10d in June 2020, and **(c)** the corresponding vegetation response (quantify by ΔSIF, units: $W \cdot m^{-2} \cdot \mu m^{-1} \cdot sr^{-1}$) over the Yangtze River Basin. **(d–e)** Same as **(b–c)** except for the IHE event in August 2022. The orange curve in **(c,e)** denotes the observed average ΔSIF over the Yangtze River Basin, and the blue curve denotes the climatological ΔSIF for the same period. The green and yellow shadings denote positive and negative ΔSIF anomalies, respectively.

## Increases in frequency and spatial extent of growing-season IHE events

During 2000–2024, there are significant increases in the occurrence of IHE events during the growing season in China. The climatology of IHE events shows broad inter-event and inter-annual variability (Supplementary Text S1 and Fig. S1). Specifically, in terms of occurrence, IHE event counts have significantly increased over the past 25 years, with a notable sharp rise after 2020. Before 2020, the frequency of IHEs (Fig. 2a) remained generally stable at 1.75 events on average per year. However, an exceptional peak emerged from 2020 to 2022, peaking at 12 events in 2021. This recent rapid rise has led to a significant upward tendency of +2.12 events per decade ($p < 0.01$) since 2000. Correspondingly, the extreme days of IHEs (Fig. 2b) also showed a rapid rising tendency, at a rate of +14.52 days per decade ($p < 0.01$).

Parallel to the increasing occurrence, the spatial extent of IHEs has also undergone a significant expansion over the past 25 years (Fig. 2c). The seasonal cumulative IHE affected area exhibits a statistically significant upward tendency of $+4.48 \times 10^6$ km$^2$ per decade ($p < 0.01$). This expansion is predominantly driven by a rapid increase in the area affected by wet zone ($+2.77 \times 10^6$ km$^2$ per decade, $p < 0.01$), with dry zone also contributing a substantial, though smaller, increase (+1.76 × $10^6$ $km^2$ per decade, $p < 0.05$).

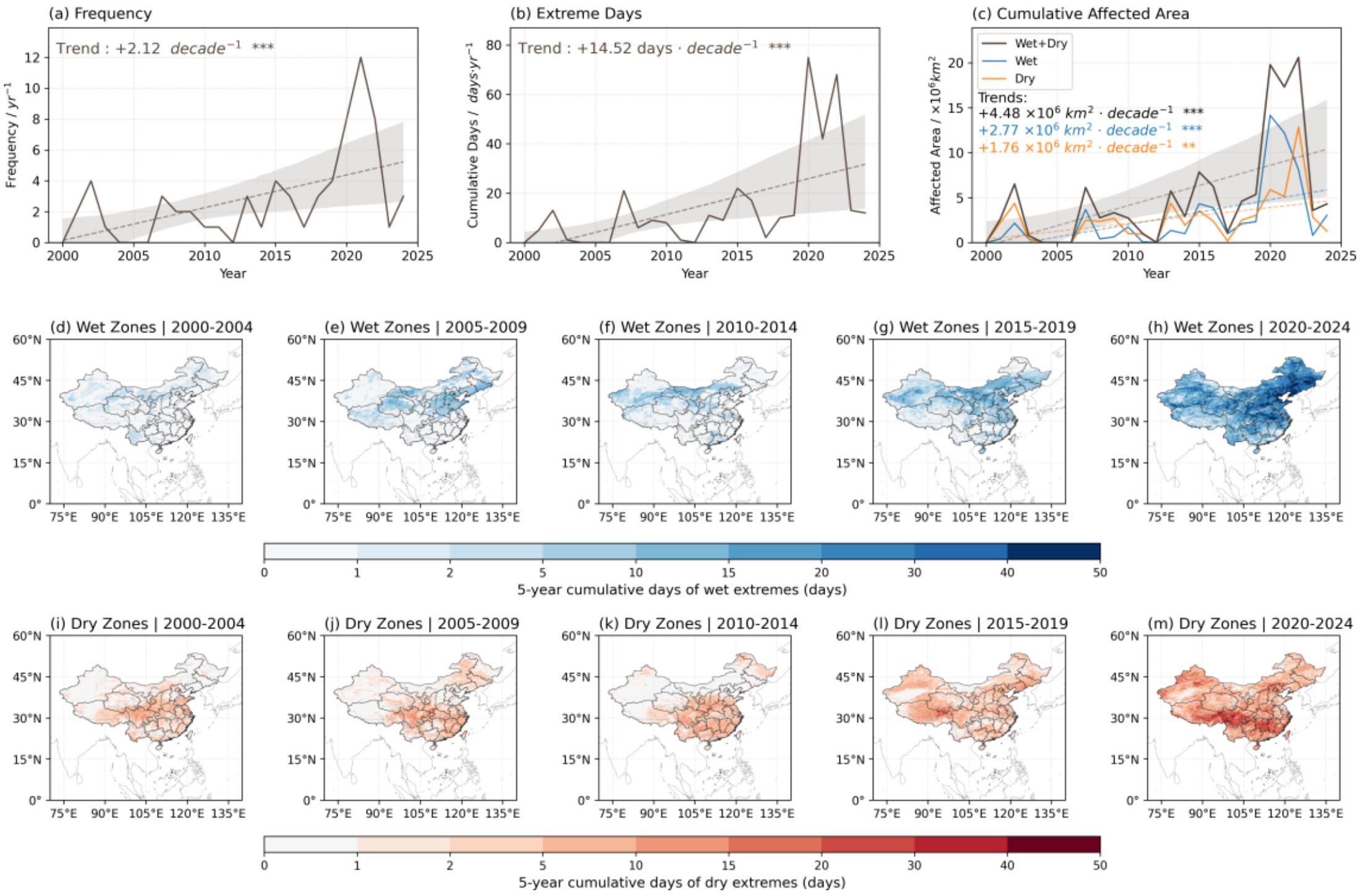

**Fig.2 | Changes in IHEs during growing season in China from 2000-2024.** Time series of **(a)** events frequency (units: counts), **(b)** extreme days (units: days), and **(c)** cumulative affected area (units: $km^2$) of IHEs. In panel **(c)**, the colored curves denote cumulative wet-zone area (blue), dry-zone area (orange), and total area (black) impacted by IHEs. Spatial patterns of 5-year cumulative occurrence days (units: days) of **(d-h)** extreme wet zones and **(i-m)** extreme dry zones in China during **(a, f)** 2000–2004, **(b, g)** 2005–2009, **(c, h)** 2010–2014, **(d, i)** 2015–2019, and **(e, j)** 2020–2024.

The notable increase in IHE-affected areas indicates more regions suffering extreme rainfall and drought simultaneously (Fig. 2c). Since the vulnerability of vegetation to precipitation and drought vary across regions, depending on the vegetation type, soil conditions, etc., it is important to investigate which part of China is exposing to the increasing risks of IHEs. To this end, for each extreme IHE day, grid points were classified as extreme wet zones (SPI > 2) and extreme dry zones (SPI < −2), and the frequency of extreme wet and dry occurrences was counted at each grid point across all extreme IHE days.

The cumulative spatial pattern suggests that most regions across China have been affected by IHEs to some extent (Supplementary Text S1 and Fig. S2). From a temporal perspective, both extreme dry and wet zones have expanded from their initial hotspots to extensive areas of China in the past 25 years. In the first half of the study period (2000–2014), the extreme wet zones of IHEs are confined to northern China (Fig. 2d,e,f), while the extreme dry zones primarily affect southwestern China (Fig. 2i,j,k), with affected areas smaller than half of the country. This distribution exhibits a quick shift in the past decade (2015–2024), characterized by a rapid expansion in the spatial extent and a significant increase in the total affected area of both extreme wet and dry zones. Specifically, the extreme wet zones of IHEs spread to the northwestern, northeastern, and eastern China (Fig. 2g,h). Meanwhile, the extreme dry zones greatly expand to the eastern Tibetan Plateau and across the Yangtze River basin (Fig. 2l,m), with slower expansion rate than that in extreme wet zones.

The above findings highlight that the more active IHEs have affected nearly every corner of China in the past 10 years, a phenomenon that was not observed at the beginning of the 21st century. These pronounced changes motivate a closer examination of the underlying physical drivers leading to the more active IHEs.

**Underlying mechanisms**

To explore the physical mechanism underlying the spatiotemporal changes in IHEs, we examine the variability of spatial inhomogeneity in synoptic-scale thermodynamic and dynamic conditions related to precipitation, as well as the changes in the climatological mean state.

Firstly, from the aspect of synoptic variability, we developed a vertical velocity–moisture coupling index (VMCI; see Methods) to characterize spatially coherent coupling between dynamic and thermodynamic conditions. In general, positive ("wet + ascent") and negative ("dry + subsidence") VMCIs correspond to the enhancement and suppression of SPI anomalies, respectively. Thus, larger absolute VMCI values are associated with stronger hydrological spatial inhomogeneity, which might evolve into IHEs.

The SV of VMCI ($SV_{VMCI}$, Supplementary Fig. S3) shows that the spatial inhomogeneity of this thermodynamic–dynamic coupling has increased significantly over the past 25 years, consistent with that in $SV_{SPI}$. After detrending, the $SV_{VMCI}$ remains highly correlated with $SV_{SPI}$ ($r=0.75$, $p < 0.01$), indicating a close spatial correspondence between SPI and VMCI, and supporting the use of VMCI to explain changes in IHE activity.

To further explain the evolution of IHE frequency and spatial extent, we examined the long-term changes in inhomogeneous VMCI extreme (IVE) events during the growing season (Fig. 3a-b). The results show that the frequency of IVEs increased by 1.2 events per decade ($p < 0.1$), the extreme days increased by 5.7 days per decade ($p < 0.1$), and the extreme affected area increased by $3 \times 10^6 km^2$ per decade ($p < 0.05$), all with a rapid rise since 2020. These changes are highly consistent with those in IHE events (Fig. 2a-b), indicating that the increasing occurrence and intensification of IHEs can be explained by those of IVEs, which have provided more favored thermodynamic and dynamic conditions for triggering IHEs. Further analysis suggests that the increasing IVE events are mainly contributed by the more frequent extremely inhomogeneous atmospheric moisture distribution (Supplementary Fig. S4).

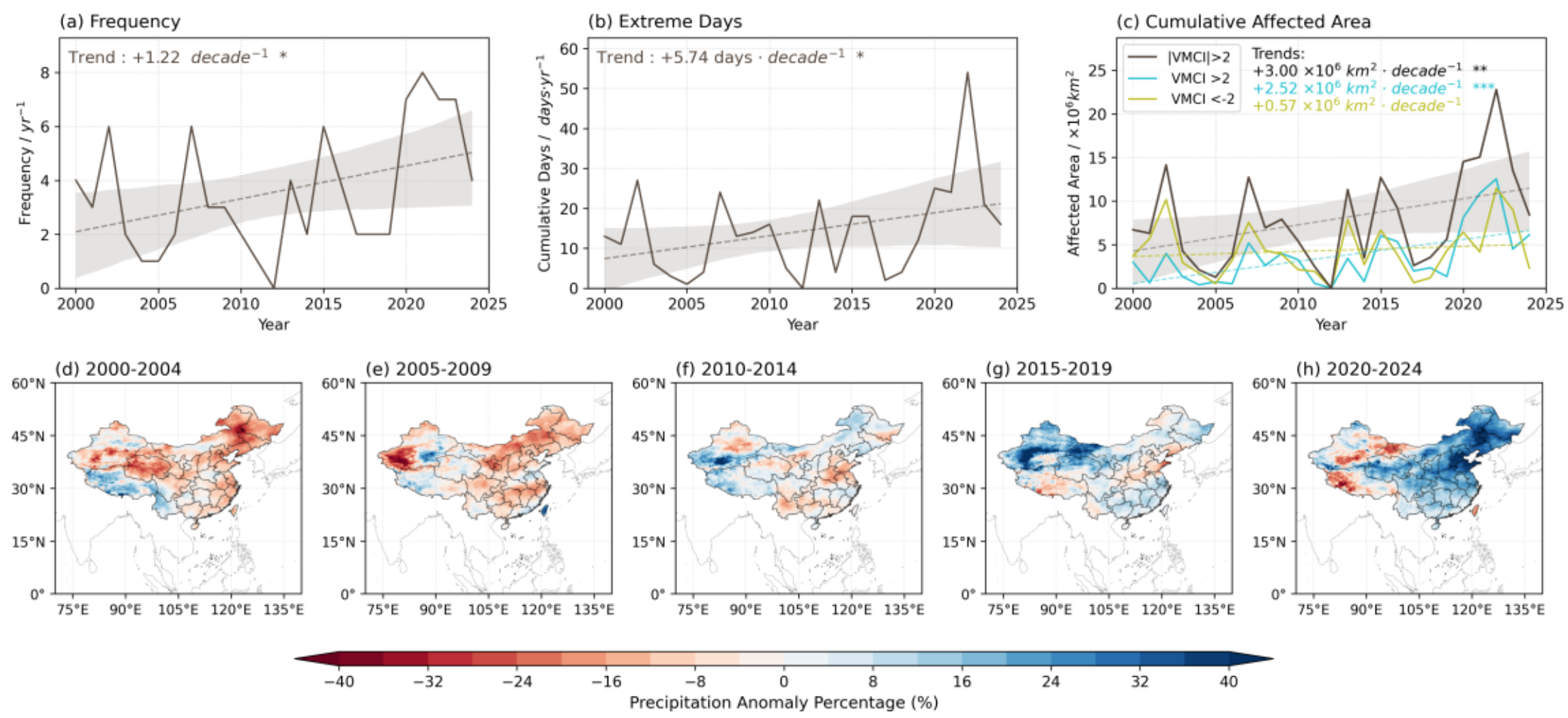

**Fig.3 | Changes in IVEs during growing season in China from 2000-2024. (a)** events frequency (units: counts), **(b)** extreme days (units: days), and **(c)** cumulative affected area (units: $km^2$). In panel **(c)**, the colored lines denote for cumulative positive VMCI area (cyan), negative VMCI area (olive), and total area (black) impacted by IVEs each year, respectively. **(d-h)** Spatial patterns of precipitation anomaly percentage (units: %) during **(d)** 2000–2004, **(e)** 2005–2009, **(f)** 2010–2014, **(g)** 2015–2019, and **(h)** 2020–2024.

However, although the intensification of IVE explains the increasing IHE events, it is not sufficient to account for the recent shifts in the IHE affected area (Fig. 2d-m). Further analysis shows that the shift in climatological precipitation distribution is the major reason modulating the spatial evolution of the IHE affected area. Over the past 25 years, the precipitation pattern exhibits an overall wetting tendency across China, with a more pronounced wetting signal emerging in northern China during the most recent five years (Fig. 3). This indicates a northward enhancement in the climatological distribution, which has been attributed to intensified moisture transport driven by the westward extension of the west Pacific subtropical high and the northward advance of the East Asian summer monsoon circulation[36–38]. Under the background, there is a higher probability for positive VMCI anomalies to occur and be translated into wet anomalies in northern China (Fig. 3d-h). By contrast, in western and southern regions where background precipitation changes are weaker, negative VMCI anomalies tend to suppress precipitation and reinforce dry anomalies.

Overall, the spatiotemporal evolution of IHEs since the beginning of the 21st century can be understood as the combined outcome of physical mechanisms at two levels. First, the strengthening of VMCI spatial heterogeneity provides the necessary thermodynamic and dynamic coupled conditions for the more frequent occurrence of IHEs. As the inhomogeneity in the coupling between vertical velocity and moisture increases, it is more likely to observe persistent and extensive coexistence of “precipitation-favorable” and “drought-favorable” regions across China. Second, the northward enhancement of the precipitation climatological background shapes regional differences in the response to hydrological anomalies. The overall northward enhance shift in background rainfall mean state increases the efficiency with which positive VMCI anomalies are translated into extreme precipitation and favoring a northward expansion of extreme wet areas in northern China, with extreme dry areas expanding in southern China. The superposition of these two effects has jointly led to the rapid increase and expansion of IHEs in recent years, and shape the “Wet-North and Dry-South” dipole-like spatial pattern in IHE-affected area (Fig. 2h,m and Supplementary Fig. S2).

**Impacts on Vegetation Growth in China**

The increasing activity of IHEs introduces compound climate risks to China. This study focuses on one of the most critical potential consequences: vegetation stress. The impacts of IHEs on vegetation are quantified using a cumulative response intensity (CRI) index derived from SIF, where positive and negative values respectively indicate vegetation growth enhancement and stress (see Methods).

Since 2000, the vegetation responses in IHE wet zones have intensified and become more regionally distinct (Fig. 4a-e). Before 2014, vegetation responses in these wet zones were relatively weak, characterized by mixed signals of stress and enhancement. In contrast, the responses have significantly strengthened over the past decade (2015–2024). Nearly all grid cells showed an increase in both positive and negative CRI values over time, indicating a growing regional discrepancy in vegetation responses to IHE wet zones. For instance, eastern Henan, northeastern Sichuan, and Inner Mongolia have emerged as recurring stress hotspots, while southwestern China consistently exhibits signals of vegetation growth enhancement. This sharpening regional differentiation demonstrates that the expanded and intensified IHE wet zones have amplified ecological responses in a geographically divergent manner.

Meanwhile, vegetation responses in IHE dry zones have worsened consistently across regions since 2000 (Fig. 4f-j). In the early 21st century, mixed signals of vegetation growth enhancement and stress were observed between the Yellow River and Yangtze River basins. However, the positive CRI areas gradually diminished, while negative responses expanded substantially to Northeast China, Northwest China, and most parts of southern China. The dominant pattern is consolidated and intensifying vegetation stress across most affected regions. This indicates that drought extremes associated with IHEs have been threatening vegetation growth in China, with conditions worsening steadily over the past two decades.

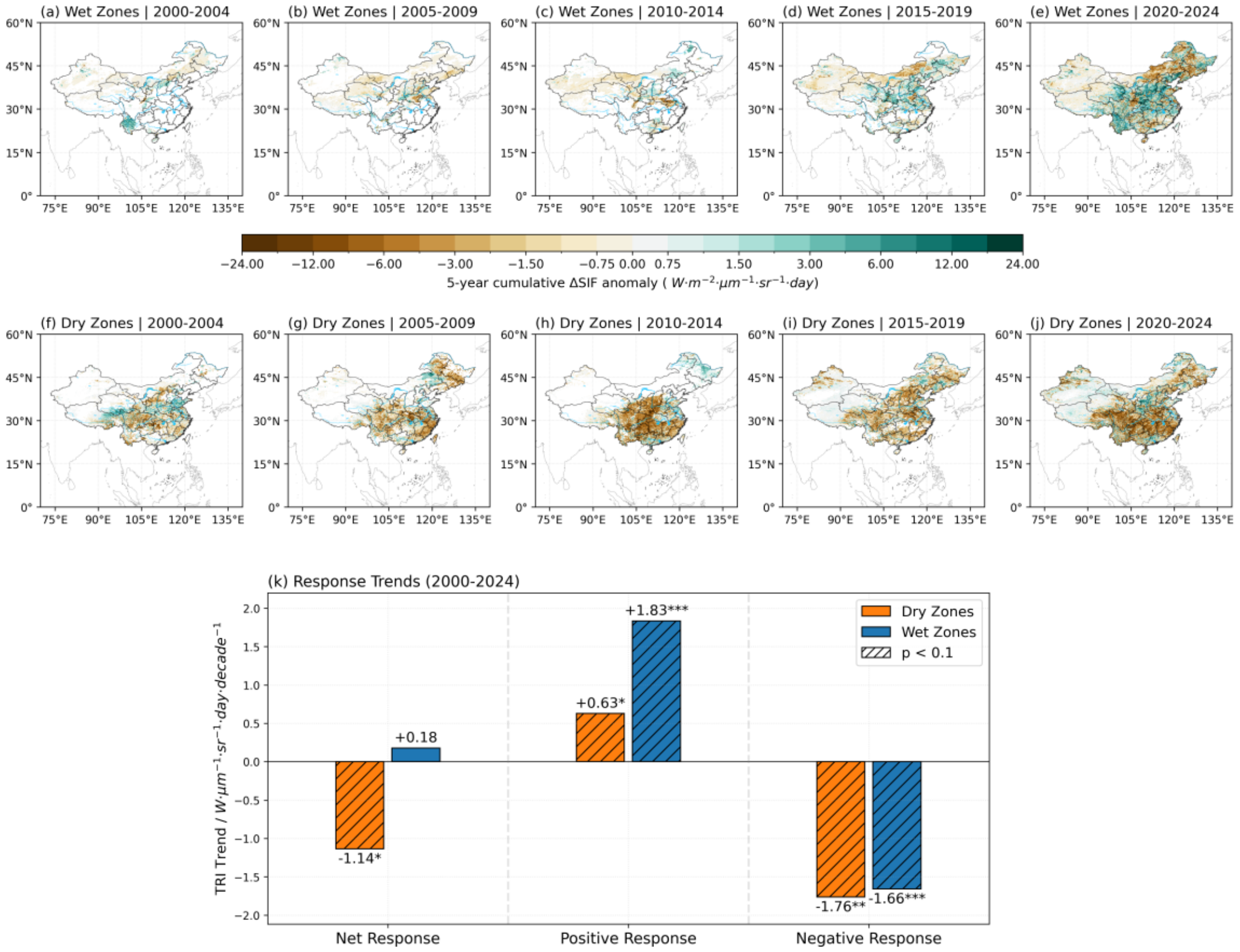

**Fig.4 | Vegetation responses to IHEs in China from 2000 to 2024. (a–j)** Spatial pattern of 5-year cumulative CRI of vegetation (units: W·m$^{-2}$·μm$^{-1}$·sr$^{-1}$·day) in **(a–e)** wet zones and **(f–j)** dry zones during IHEs in China for **(a, f)** 2000–2004, **(b, g)** 2005–2009, **(c, h)** 2010–2014, **(d, i)** 2015–2019, and **(e, j)** 2020–2024, respectively. (k) Linear trends in TRI (units: W·μm$^{-1}$·sr$^{-1}$·day·decade$^{-1}$) over 2000–2024 for net, positive, and negative responses in wet and dry zones. Net response is defined as the sum of positive and negative responses. Hatched bars indicate statistical significance at $p < 0.1$.

These results indicate that while the impacts of IHE wet extremes are spatially diverse, vegetation stress induced by IHE droughts has intensified persistently and become predominant. This contrast is evident in the total response index (TRI, see Methods), which quantifies the overall vegetation response to IHEs across China. The annual time series of TRI shows the annual evolution of these responses (Supplementary Fig. S5), whereas Fig. 4k summarizes their linear trends. Over the 25-year study period, the TRI values for both vegetation growth enhancement and stress in IHE wet zones enhanced significantly, and these two opposing responses offset each other, leading to a relatively stable net TRI (Fig. 4k). In contrast, in dry zones, the TRI change for vegetation stress was larger than that for growth enhancement, resulting in a strong and persistent negative tendency in the net TRI (Fig. 4k). This decline accelerated dramatically over the past five years, as multiple IHE events set new records for the severity of vegetation stress [27,39].

Overall, the spatial pattern of changing IHE activity in recent years poses a unique ecological threat to China. The coexistence of extreme wet conditions adjacent to drought-affected areas does not necessarily facilitate vegetation recovery. Instead, the widespread vegetation stress in dry zones may eliminate potential ecological refuges that are critical for post-disturbance regeneration. This large-scale synchronized impairment disrupts metacommunity dynamics and landscape-scale recovery pathways, thereby eroding ecosystem resilience and potentially leading to long-term degradation. Therefore, the increasing frequency of IHEs exposes ecosystems to recurrent, landscape-scale compound hydrological shocks, which may directly amplify the scale and intensity of vegetation stress across China.

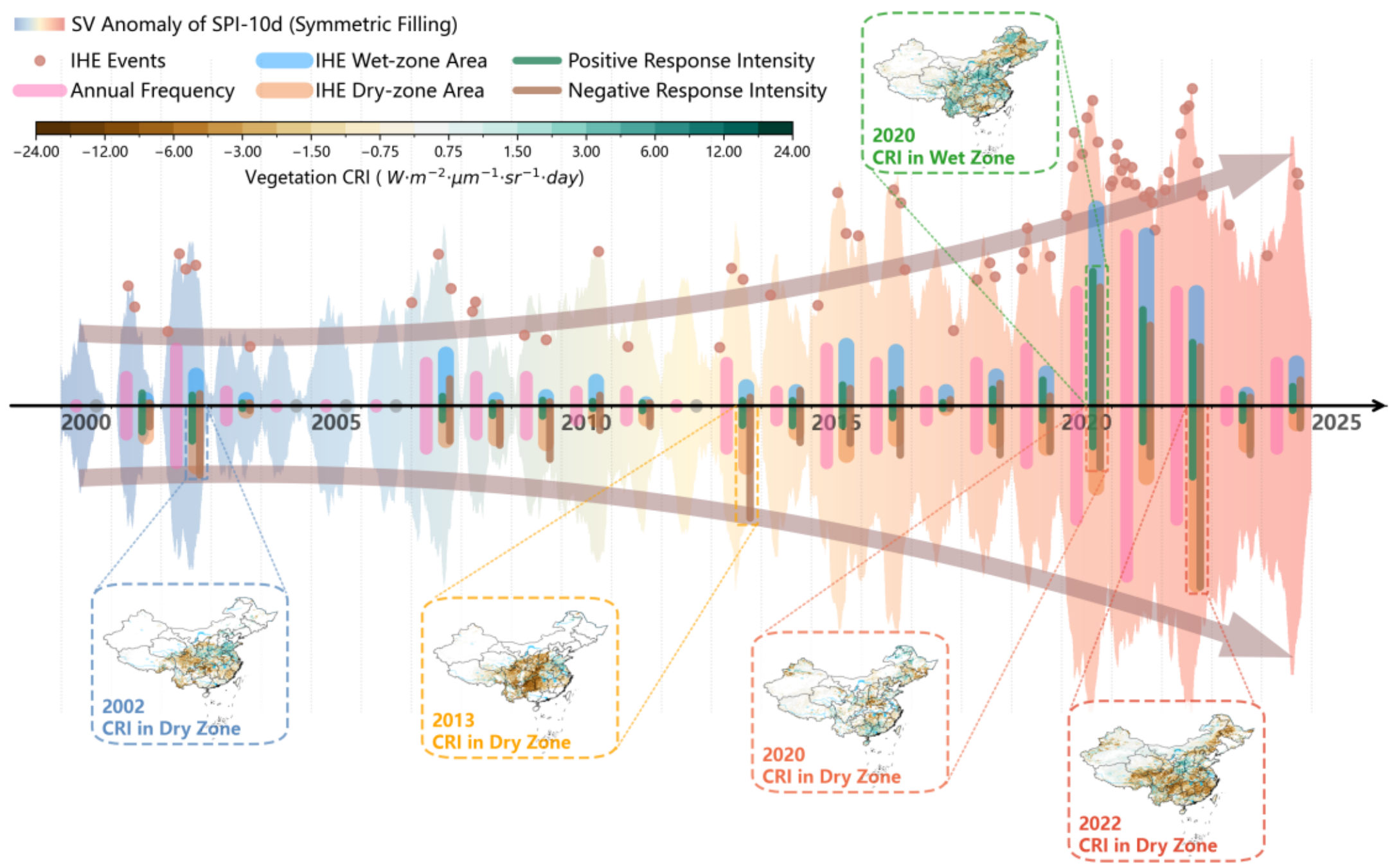

**Fig.5 | Schematic diagram showing the rapid increase in inhomogeneous hydrological extremes stress vegetation growth in China.** Background shading shows the SV anomaly of SPI-10d during the growing season from 2000 to 2024. Dots denote all identified IHE events. Pink bars indicate annual event frequency. Blue and orange bars indicate the area affected by wet-zone and dry-zone during IHE events, respectively. Green and brown bars indicate positive and negative vegetation response intensity in wet-zone (above the x-axis) and dry-zone (below the x-axis), respectively. Grey arrows schematically indicate the recent acceleration in both the extent of IHEs and the intensity of vegetation responses. Insets show representative spatial patterns of vegetation CRI during selected years dominated by dry-zone impacts (2002, 2013, 2020 and 2022) and wet-zone impacts (2020).

## Discussion

IHEs are closely related to compound and concurrent extremes. Compound extreme events involve the co-occurrence of two or more climate drivers or hazards, emphasizing the diversity of hazard types that interact locally[40,41]. Concurrent extremes, referring to the simultaneous occurrence of similar or different extreme events across distant regions, have gained increasing attention due to their profound implications for food security, global supply chains, and disaster management[7,42]. IHEs can be regarded as a kind of concurrent extremes in which different subregions are simultaneously exposed to impactful but contrasting phases of the same hazard (e.g., coexisting droughts and floods). Such events can exert substantial pressures on the ecological and socio-economic systems[43].

In this study, we find that the enhancing precipitation inhomogeneity has led to a significant rise in inhomogeneous hydrological condition activity during the growing season. Since 2000,

IHEs in China have experienced a significant increase in frequency, with a sharp rise in the past 5 years, and rapidly expanded from regional anomalies to a national-scale hazard (Fig. 5). Driven by the intensifying inhomogeneity of moisture-dynamic coupled conditions and the northward migration of the precipitation climatological distribution, the affected area of IHEs has gradually evolved to a distinct “Wet-North and Dry-South” pattern.

It should be noted that IHEs do not respond linearly to anomalies in thermodynamic–dynamic coupling, but are modulated by additional factors. For example, strong thermodynamic–dynamic coupling may sometimes remain below the extreme threshold, but prolonged accumulation can still induce IHEs. Other factors may also amplify the precipitation response under recent climate change, such as increased atmospheric stability which suppresses weak precipitation but intensifies heavy rainfall once convection initiates. Nevertheless, the highly consistent tendencies and variability between IHEs and IVEs confirms that VMCI changes are a primary driver of enhanced IHE activity.

With the rising frequency of IHEs, simultaneous exposure to severe dry and wet conditions across diverse regions increases systemic ecological risks on a national scale. Our analysis demonstrates that the expansion of drought-prone areas exerts stronger and more consistent stress on vegetation than the compensatory effects of heavy precipitation, underscoring the asymmetric ecological consequences of precipitation inhomogeneity (Fig. 5). This exacerbated stress is attributed to both plant physiological mechanisms and the unique spatial configuration of IHEs. Firstly, drought stress may push plant hydraulic systems closer to their failure thresholds[13,39,44]. Worse still, the large-scale and synchronized nature of these IHE dry zones amplifies the ecological impact beyond the physiological scale. When extreme drought synchronously affects vast, contiguous areas it can overwhelm regional climatic buffering capacity, triggering abrupt, large-scale declines in vegetation productivity. An example is the 2023 record-breaking productivity collapse[39] in southwestern China, a hotspot for IHE dry extremes (Fig. 2) and intense vegetation stress responses (Fig. 4).

More active IHEs can threaten national food security and terrestrial ecosystems viability. On an agricultural level, IHEs can disrupt national agricultural stability, making it difficult to offset yield losses in one region with surpluses elsewhere. Beyond agriculture, the large-scale afforestation and ecological restoration projects, which have driven global greening[30] and improved regional water cycling in China's arid and semi-arid regions[31], are vulnerable to IHE-induced hydrological stresses. Over the past two decades, nearly half of China’s forest ecosystems have experienced declining resilience, mainly due to intensified droughts and increased precipitation variability[45]. Our results further reveal growing vegetation stress from IHEs in northwestern China and Inner Mongolia (Fig. 4e & 5). While rainfall in IHEs can temporarily alleviate drought conditions, its concentrated inputs exceed vegetation uptake and soil infiltration capacity, thereby intensifying surface runoff and erosion, and even inducing root anoxia and physical damage, which ultimately suppresses normal plant growth. Looking forward, expanding IHEs will likely expose vegetation to more complex, uncertain hydroclimatic conditions, characterized by alternating droughts and floods, amplified soil moisture fluctuations, and reduced synchrony between water demand and supply. These findings highlight the need for future vegetation management to prioritize system resilience and adaptability, strengthening region-specific adaptation to complex hydroclimatic regimes, thereby ensuring long-term sustainability.

## Methods

### Data

The precipitation dataset used in this study is derived from the Global Precipitation Measurement (GPM) mission, specifically the IMERG V7 L3 daily precipitation product (GPM_3IMERGDF) covering the period 2000–2024. GPM is an international satellite mission jointly led by National Aeronautics and Space Administration (NASA) and Japan Aerospace Exploration Agency (JAXA), which integrates multiple sensors, satellites, and algorithms with ground-based rain gauge data to provide high-accuracy precipitation estimates at a spatial resolution of 0.1°×0.1°.

To capture precipitation variability at scales relevant to extreme events and their hydrological impacts, we employ the standardized precipitation index at a 10-day timescale (SPI-10d) as the precipitation metric. The 10-day SPI is used to avoid the overweighting of wet extremes during the SV calculation, while keeping the timescale matching synoptic-scale events and their immediate impacts on vegetation. Each SPI value is calculated from the precipitation distribution of the 21-day moving window centered on the same calendar day over the 25-year period (i.e., a set of 21 days×25 years=525 samples). This wider sampling window ensures robust fitting in regions and seasons with sparse rainfall while retaining sensitivity to sub-seasonal extremes. For consistency, all SPI values at each grid are constrained within the range [−3.1,3.1].

To evaluate the sensitivity of our results to dataset choice, we have compared the results obtained from eight precipitation datasets, including GPM IMERG V7B[46], GPCP V3.3[47], ERA5[48], CHM V2.1[49], CMORPH V1.0[50], PERSIANN-CDR V1R1[51], CN05.1[52], and GPCC V2022[53] (see Supplementary Fig. S6 and Table S1). As the results were generally consistent across datasets, GPM was selected as the representative dataset for the study.

The specific humidity and vertical velocity used in the study were obtained from the ERA5 reanalysis dataset[48] for 2000–2024. ERA5 is the fifth-generation atmospheric reanalysis produced by the European Centre for Medium-Range Weather Forecasts (ECMWF), providing globally complete hourly atmospheric fields at 0.25° × 0.25° spatial resolution.

Vegetation activity is quantified using SIF, the faint emission in the 650–800 nm range released from chloroplasts during photosynthesis. SIF is tightly coupled with photosynthetic electron transport and expressed in units of $W \cdot m^{-2} \cdot \mu m^{-1} \cdot sr^{-1}$. While greenness-based indices, such as NDVI, which primarily reflect canopy morphology, SIF provides a direct and more immediate proxy for photosynthetic function, offering improved sensitivity in both evergreen and deciduous systems. Its correlation with gross primary production (GPP) is consistently stronger than that of traditional vegetation indices, making SIF a powerful indicator of vegetation stress and productivity, particularly under short-term extreme weather conditions. We employ the GOSIF v2.0 dataset[54], offering global 0.05° SIF fields at 8-day intervals for 2000–2024, derived from the Orbiting Carbon Observatory-2 (OCO-2) observations via machine-learning method.

### Definition of Spatial Inhomogeneity

Spatial inhomogeneity refers to the variability of a variable across space and is commonly quantified using the spatial standard deviation (SSD)[55] or SV[6], which measure the absolute dispersion in space. For any variable $A_{(x,y,t)}$, its SV at time *t* is defined as:

$$SV_A(t) = \frac{\sum_{x,y}\left[A_{(x,y,t)} - \overline{A_{(t)}}\right]^2 S_{(x,y)}}{\sum_{x,y} S_{(x,y)}} \quad (1)$$

where $x, y$ denote grid point in latitude and longitude coordinates, $t$ represents time, $S_{(x,y)}$ is the area of each grid, which is used as the weighting factor in averaging, and $\Sigma_{x,y}$ denotes the spatial summation over all grid points. $\overline{A_{(t)}}$ represents the area-weighted spatial mean of $A$ at time $t$.

Compared with the spatial mean, SV, derived from the second moment, gives greater weight to the tails of the probability distribution and is thus more sensitive to the concurrence of strong positive and negative anomalies. This makes it particularly suitable for detecting spatially inhomogeneous hydroclimatic conditions, such as the simultaneous occurrence of wet and dry anomalies in different parts of the region. For SPI, which is standardized relative to the local climatology, the regional mean is generally close to zero. Thus, to quantify the dispersion of spatial wet–dry departures from the climatology, we approximate $\overline{SPI_{(t)}} \equiv 0$, and Eq.1 is simplified to:

$$SV_{SPI}(t) = \frac{\sum_{x,y} SPI_{(x,y,t)}{}^{2} S_{(x,y)}}{\sum_{x,y} S_{(x,y)}} \tag{2}$$

**Definition of standardized thermodynamic and dynamic anomalies and vertical velocity-moisture coupling index (VMCI)**

To diagnose the spatial inhomogeneity of thermodynamic and dynamic conditions related to IHEs, we define standardized anomalies of moisture and vertical motion at the 10-day scale. Specifically, $Z_q$ and $Z_\omega$ denote the normalized anomalies of 500-hPa specific humidity ($q$) and 500-hPa vertical velocity (ω), respectively.

Because both $q$ and ω may exhibit skewed distributions, their 10-day time series for each calendar day were fitted using a Gamma distribution, followed by an equiprobability transformation to the standard normal distribution, analogous to the computation of SPI. For ω, a sign reversal was first applied so that positive (negative) values represent anomalous upward (downward) motion. To satisfy the positivity requirement of Gamma fitting, the sign-reversed ω series was shifted by a small constant offset defined as ($offset = 0.001 \times std - min$), where $std$ and $min$ are the standard deviation and the minimum of the multi-year series for a given calendar day, respectively. The shifted series is then ($\omega_{\text{shift}} = \omega + offset$). This process ensures that all samples are positive for Gamma fitting while preserving the variance, magnitude, and relative distributional characteristics of the original series.

Based on $Z_q$ and $Z_\omega$, we construct a new diagnostic metric, the vertical velocity–moisture coupling index (VMCI), defined as

$$\text{VMCI} = \begin{cases} Z_q \times Z_\omega, & if\ Z_q > 0\ and\ Z_\omega > 0 \\ -(Z_q \times Z_\omega), & if\ Z_q < 0\ and\ Z_\omega < 0 \\ 0, & if\ (Z_q \times Z_\omega) \le 0 \end{cases} \tag{3}$$

A positive VMCI indicates anomalously moist conditions accompanied by anomalous upward motion, which favor convection and precipitation. A negative VMCI indicates anomalously dry conditions accompanied by anomalous downward motion, corresponding to strong suppression of convection and drying conditions. When $Z_q$ and $Z_\omega$ have opposite signs, moisture and dynamic conditions are regarded as uncoupled, and VMCI is set to zero. Hence, the VMCI captures two coupled states that are directly relevant to the formation and suppression of precipitation, namely “wet + ascent” and “dry + subsidence.” It is therefore better suited to diagnosing the processes that drive the synchronous intensification of the spatial manifestation of hydrological extremes.

**Definition of inhomogeneous extremes**

In this study, inhomogeneous extremes are defined and identified using the daily $SV_V$ based on two primary criteria (as diagram shown in Supplementary Fig. S7):

1. Duration ≥ 1 day: The daily $SV_V$ exceeded the seasonal 95th percentile for at least one day.
2. Interrupt ≤ 3 days: Successive exceedances separated by ≤3 days were merged into a single event.

Inhomogeneous extremes are identified in a unified way for all standardized variables. Specifically, the extreme events identified from $SV_{SPI}$ are termed inhomogeneous hydrological extremes (IHEs), and the extreme events identified from $SV_{VMCI}$ are termed inhomogeneous VMCI extremes (IVEs).

Given the pronounced seasonality of precipitation across the monsoon-dominated study region, and the fact that inhomogeneous wet/dry extremes mainly occur during the active growing season, winter (December–February) is excluded from the event analysis. We therefore focus on the growing season from March to November.

**Event metrics and affected area**

For each type of inhomogeneous extremes, we derive three annual metrics: occurring frequency, cumulative extreme days, and cumulative affected area. Event frequency is defined as the number of events within the growing season per year, and cumulative extreme days is the annual sum of days with events being observed.

The affected area is defined based on the corresponding standardized variables itself. For a given standardized variable $X$ (i.e., $SPI$, VMCI, $Z_q$ and $Z_\omega$), we first calculate the temporal maximum and minimum of $X$ at each grid cell during the event period, denoted by $X_{max}$ and $X_{min}$, respectively. Grid cells satisfying $X_{max} > 2$ or $X_{min} < -2$ during the event days are regarded as affected areas under strong positive or negative anomalies, respectively. The total affected area of that event is then defined as the sum of these two areas. Because all variables were transformed into standardized anomaly space, the threshold of ±2 provides a consistent criterion for identifying strong local anomalies, approximately corresponding to the 5th and 95th percentiles of the data points.

Based on this definition, the seasonal cumulative affected area is calculated as the sum of the total affected areas of all events occurring within the growing season of a given year. If the same grid cell is affected by multiple events in the same year, it is counted repeatedly.

For example, in IHEs, the affected area is specifically defined based on SPI-10d. Since SPI-10d is primarily designed to quantify precipitation anomalies, with larger absolute values indicating stronger departures from climatology, thresholds of ±2 are commonly used to represent extreme conditions, roughly corresponding to the 5th and 95th percentiles of the precipitation distribution. Accordingly, regions with SPI-10d > 2 (wet zones) or SPI-10d < –2 (dry zones) during IHEs are defined as the affected areas.

**Quantifying anomaly in vegetation response**

Ecosystem resilience causes vegetation responses to extreme drought or precipitation to be lagged. To capture the full impact, we quantified the cumulative post-event anomalies in vegetation:

**Step 1**: For each time point after an event, calculate the difference in vegetation index $\Delta SIF$ between the post-event period and the pre-event:

$$\Delta SIF_{(id,x,y,lag\ time)} = SIF_{post-event(id,x,y,lag\ time)} - SIF_{pre-event(id,x,y)},$$

**Step 2**: For each time point, calculate the anomaly of ΔSIF relative to the climatological mean $\Delta SIF$ for the corresponding calendar period, where the climatological mean is estimated from 2000–2024:

$$\Delta SIF_{anom\ (id,x,y,lag\ time)} = \Delta SIF_{(id,x,y,lag\ time)} - \overline{\Delta SIF_{multiyear}}_{(x,y,lag\ time)},$$

The $\Delta SIF_{anom}$ accounts for changes in vegetation growth after an IHE event occurs, while removing the seasonal cycle of phenology, and makes comparison with the climatological baseline.

**Step 3**: For each event, calculate the cumulative impact at each lag time, defined as cumulative response intensity (CRI):

$$CRI_{(id,x,y,lag\ time)} = \sum_{i=0}^{lag\ time} \Delta SIF_{anom\ (id,x,y,i)}.$$

We further calculate the area-weighted summation over Chinas, termed the total response intensity (TRI) of vegetation anomalies, defined as:

$$TRI_{(id,lag\ time)} = \sum_{x,y} CRI_{(id,x,y,lag\ time)} S_{(x,y)}$$

In this study, we set the lag time to 64 days, capturing the majority of vegetation's delayed responses while minimizing potential overlap with subsequent events that could introduce artifacts.

**References**

1. Hsu, H. & Fueglistaler, S. Robust Projections of Changing Precipitation Evenness in a Warming Climate. *Geophys. Res. Lett.* **52**, e2025GL114953 (2025).
2. Ghanghas, A., Sharma, A., Dey, S. & Merwade, V. How Is Spatial Homogeneity in Precipitation Extremes Changing Globally? *Geophys. Res. Lett.* **50**, e2023GL103233 (2023).
3. Hervé Douville *et al.* Water Cycle Changes. in *Climate Change 2021: The Physical Science Basis. Contribution of Working Group I to the Sixth Assessment Report of the Intergovernmental Panel on Climate Change [Masson-Delmotte, V., P. Zhai, A. Pirani, S.L. Connors, C. Péan, S. Berger, N. Caud, Y. Chen, L. Goldfarb, M.I. Gomis, M. Huang, K. Leitzell, E. Lonnoy, J.B.R. Matthews, T.K. Maycock, T. Waterfield, O. Yelekçi, R. Yu, and B. Zhou (eds.)]* 1055–1210 (Cambridge University Press, Cambridge, United Kingdom and New York, NY, USA, 2021).
4. Zhang, W. *et al.* Increasing precipitation variability on daily-to-multiyear time scales in a warmer world. *Sci. Adv.* **7**, eabf8021 (2021).
5. Zhang, Y. & Fueglistaler, S. Mechanism for Increasing Tropical Rainfall Unevenness With Global Warming. *Geophys. Res. Lett.* **46**, 14836–14843 (2019).
6. Liu, S. *et al.* A general framework quantifying variability in spatial inhomogeneity of global precipitation and its contribution. *Clim. Dyn.* **63**, 129 (2025).
7. Zhou, S., Yu, B. & Zhang, Y. Global concurrent climate extremes exacerbated by anthropogenic climate change. *Sci. Adv.* **9**, eabo1638 (2023).
8. Gui, Y. *et al.* Vegetation greenness in 2024. *Nat. Rev. Earth Environ.* **6**, 255–257 (2025).
9. Chen, X. *et al.* The global greening continues despite increased drought stress since 2000.

*Glob. Ecol. Conserv.* **49**, e02791 (2024).
10. Wei, S., Li, X., Wang, K., Wang, T. & Piao, S. Two decades of persistent greening in China despite 2023 climate extremes. *Sci. China Earth Sci.* **68**, 1064–1073 (2025).
11. Yan, Y. *et al.* Climate-induced tree-mortality pulses are obscured by broad-scale and long-term greening. *Nat. Ecol. Evol.* **8**, 912–923 (2024).
12. Lesk, C., Rowhani, P. & Ramankutty, N. Influence of extreme weather disasters on global crop production. *Nature* **529**, 84–87 (2016).
13. Ohlert, T. *et al.* Drought intensity and duration interact to magnify losses in primary productivity. *Science* **390**, 284–289 (2025).
14. Zhou, S., Zhang, Y., Park Williams, A. & Gentine, P. Projected increases in intensity, frequency, and terrestrial carbon costs of compound drought and aridity events. *Sci. Adv.* **5**, eaau5740 (2019).
15. Bai, Y., Chen, J., Zhang, Y. & Tang, Z. Response Modes of Global Vegetation to Extreme Drought. *Glob. Change Biol.* **31**, e70488 (2025).
16. Doughty, C. E. *et al.* Drought impact on forest carbon dynamics and fluxes in Amazonia. *Nature* **519**, 78–82 (2015).
17. Anderegg, W. R. L. *et al.* Pervasive drought legacies in forest ecosystems and their implications for carbon cycle models. *Science* **349**, 528–532 (2015).
18. Choat, B. *et al.* Triggers of tree mortality under drought. *Nature* **558**, 531–539 (2018).
19. Wood, R. R. Role of mean and variability change in changes in European annual and seasonal extreme precipitation events. *Earth Syst. Dyn.* **14**, 797–816 (2023).
20. Saharia, M. *et al.* On the Impact of Rainfall Spatial Variability, Geomorphology, and Climatology on Flash Floods. *Water Resour. Res.* **57**, e2020WR029124 (2021).
21. Liu, J. *et al.* Hydrological effects of climate variability and vegetation dynamics on annual fluvial water balance in global large river basins. *Hydrol. Earth Syst. Sci.* **22**, 4047–4060 (2018).
22. Huang, W. K., Monahan, A. H. & Zwiers, F. W. Estimating concurrent climate extremes: A conditional approach. *Weather Clim. Extrem.* **33**, 100332 (2021).
23. Jiang, Q. *et al.* Complex Networks Reveal Climate Models' Capability in Simulating Global Synchronized Extreme Precipitation. *Geophys. Res. Lett.* **53**, e2025GL118219 (2026).
24. Cai, Y., Chen, Z. & Du, Y. The role of Indian Ocean warming on extreme rainfall in central China during early summer 2020: without significant El Niño influence. *Clim. Dyn.* **59**, 951–960 (2022).
25. Zhou, Z.-Q., Xie, S.-P. & Zhang, R. Historic Yangtze flooding of 2020 tied to extreme Indian Ocean conditions. *Proc. Natl. Acad. Sci.* **118**, e2022255118 (2021).
26. Tan, X. *et al.* Detection and attribution of the decreasing precipitation and extreme drought 2020 in southeastern China. *J. Hydrol.* **610**, 127996 (2022).
27. Wang, J. *et al.* Unprecedented decline in photosynthesis caused by summer 2022 record-breaking compound drought-heatwave over Yangtze River Basin. *Sci. Bull.* **68**, 2160–2163 (2023).
28. Yuan, X., Wang, Y., Zhou, S., Li, H. & Li, C. Multiscale causes of the 2022 Yangtze mega-flash drought under climate change. *Sci. China Earth Sci.* **67**, 2649–2660 (2024).
29. Li, Z.-L. & Jiao, X.-Z. Evaluation and projections of summer daily precipitation over Northeastern China in an optimal CMIP6 Multimodel Ensemble. *Clim. Dyn.* **62**, 6235–6251

(2024).
30. Chen, C. *et al.* China and India lead in greening of the world through land-use management. *Nat. Sustain.* **2**, 122–129 (2019).
31. Li, Y. *et al.* Divergent hydrological response to large-scale afforestation and vegetation greening in China. *Sci. Adv.* **4**, eaar4182 (2018).
32. Tu, Y. *et al.* A 30 m annual cropland dataset of China from 1986 to 2021. *Earth Syst. Sci. Data* **16**, 2297–2316 (2024).
33. Liu, M., Zhou, X., Huang, G. & Li, Y. The increasing water stress projected for China could shift the agriculture and manufacturing industry geographically. *Commun. Earth Environ.* **5**, 396 (2024).
34. Zhu, Y. *et al.* Heterogeneity in Spatiotemporal Variability of High Mountain Asia's Runoff and Its Underlying Mechanisms. *Water Resour. Res.* **59**, e2022WR032721 (2023).
35. Zhang, L. *et al.* Understanding and Attribution of Extreme Heat and Drought Events in 2022: Current Situation and Future Challenges. *Adv. Atmospheric Sci.* **40**, 1941–1951 (2023).
36. Diao, Y., Guo, J., Zhang, Y., Hou, Z. & Luo, B. Trend turning of North China summer extreme precipitations around early 2000s and its possible reason. *Clim. Dyn.* **61**, 5367–5386 (2023).
37. Choi, W. & Kim, K.-Y. Summertime variability of the western North Pacific subtropical high and its synoptic influences on the East Asian weather. *Sci. Rep.* **9**, 7865 (2019).
38. Zhao, Y. *et al.* Enhancement of the summer extreme precipitation over North China by interactions between moisture convergence and topographic settings. *Clim. Dyn.* **54**, 2713–2730 (2020).
39. Wang, Z. *et al.* Synergistic effects of high atmospheric and soil dryness on record-breaking decreases in vegetation productivity over Southwest China in 2023. *Npj Clim. Atmospheric Sci.* **8**, 6 (2025).
40. Zscheischler, J. *et al.* A typology of compound weather and climate events. *Nat. Rev. Earth Environ.* **1**, 333–347 (2020).
41. Zscheischler, J. *et al.* Future climate risk from compound events. *Nat. Clim. Change* **8**, 469–477 (2018).
42. Xu, Y. *et al.* Increased Significance of Global Concurrent Hazards From 1981 to 2020. *Earths Future* **12**, e2024EF004490 (2024).
43. Worou, K. & Messori, G. Compounding droughts and floods amplify socio-economic impacts. *Environ. Res. Lett.* **20**, 104024 (2025).
44. Anderegg, W. R. L. *et al.* A climate risk analysis of Earth's forests in the 21st century. *Science* **377**, 1099–1103 (2022).
45. Wang, W. *et al.* Increasing Aridity and Interannual Precipitation Variability Drives Resilience Declines in Restored Forests Across China. *Earths Future* **13**, e2025EF006164 (2025).
46. Huffman, G., Stocker, E., Bolvin, D., Nelkin, E. & Tan, J. GPM IMERG Final Precipitation L3 1 day 0.1 degree x 0.1 degree V07. 0.000 bytes UCAR/NCAR - Research Data Archive https://doi.org/10.5065/7DE2-M746 (2024).
47. Huffman, G. J. GPCP Precipitation Level 3 Daily 0.5-Degree V3.3. NASA Goddard Earth Sciences Data and Information Services Center https://doi.org/10.5067/MEASURES/GPCP/DATA307 (2024).
48. Copernicus Climate Change Service. Complete ERA5 global atmospheric reanalysis.

Copernicus Climate Change Service (C3S) Climate Data Store (CDS) https://doi.org/10.24381/CDS.143582CF (2023).
49. Hu, J. & Miao, C. CHM_PRE V2: A new upgraded high-precision gridded precipitation dataset considering spatiotemporal and physical correlations over China. Zenodo https://doi.org/10.5281/ZENODO.14632157 (2025).
50. Pingping Xie *et al.* NOAA Climate Data Record (CDR) of CPC Morphing Technique (CMORPH) High Resolution Global Precipitation Estimates, Version 1. NOAA National Centers for Environmental Information https://doi.org/10.25921/W9VA-Q159 (2018).
51. Ashouri, H. *et al.* PERSIANN-CDR: Daily Precipitation Climate Data Record from Multisatellite Observations for Hydrological and Climate Studies. *Bull. Am. Meteorol. Soc.* **96**, 69–83 (2015).
52. Wu, J., Gao, X., Giorgi, F. & Chen, D. Changes of effective temperature and cold/hot days in late decades over China based on a high resolution gridded observation dataset. *Int. J. Climatol.* **37**, 788–800 (2017).
53. Markus, Z. *et al.* GPCC Full Data Daily Version 2022 at 1.0°: Daily Land-Surface Precipitation from Rain-Gauges built on GTS-based and Historic Data: Globally Gridded Daily Totals. approx. 25 MB per gzip file Global Precipitation Climatology Centre (GPCC, http://gpcc.dwd.de/) at Deutscher Wetterdienst https://doi.org/10.5676/DWD_GPCC/FD_D_V2022_100 (2022).
54. Li, X. & Xiao, J. A Global, 0.05-Degree Product of Solar-Induced Chlorophyll Fluorescence Derived from OCO-2, MODIS, and Reanalysis Data. *Remote Sens.* **11**, 517 (2019).
55. Ren, Q. *et al.* Increasing Inhomogeneity of the Global Oceans. *Geophys. Res. Lett.* **49**, (2022).

**Acknowledgements**

This work is supported by the National Natural Science Foundation of China (72293604), the Innovation Research Foundation of National University of Defense Technology (202402-YJRC-LJ-001), the National Natural Science Foundation of China (42405038, 42405003) and the Project of Key Laboratory of Guangdong Provincial Department of Education (2025KSYS009).We also thank for the technical support of the National Large Scientific and Technological Infrastructure "Earth System Numerical Simulation Facility" (https://cstr.cn/31134.02.EL).

**Author contributions**

S L: conceptualization, methodology, formal analysis, investigation, data curation, writing—original draft, visualization;

J C H L: conceptualization, methodology, investigation, validation, writing—review & editing, funding acquisition;

J X: conceptualization, methodology, supervision, project administration, funding acquisition, writing—review & editing;

K X: methodology, writing—review & editing;

S T: methodology, validation, data curation, writing—review & editing;

M Z: data curation, formal analysis, writing—review & editing;

D X: validation, writing—review & editing;

Y L: supervision, project administration, writing—review & editing, funding acquisition;

B Z: conceptualization, methodology, formal analysis, validation, writing—review & editing, supervision, project administration, funding acquisition.

All authors reviewed the manuscript.

**Competing Interests**

The authors declare there are no conflicts of interest for this manuscript.

**Data availability**

Data used in this study can be downloaded from the websites below:

GPM IMERG: https://disc.gsfc.nasa.gov/datasets/GPM_3IMERGDF_07/summary

GPCP: https://disc.gsfc.nasa.gov/datasets/GPCPDAY_3.3/summary

ERA5: https://cds.climate.copernicus.eu/datasets/reanalysis-era5-complete

CHM_PRE: https://zenodo.org/records/15735374

CMORPH: https://gdex.ucar.edu/datasets/d502002

PERSIANN-CDR: https://www.ncei.noaa.gov/data/precipitation-persiann/access

CN05.1: https://nzc.iap.ac.cn/content?cid=24&aid=999

GPCC: https://opendata.dwd.de/climate_environment/GPCC/html/download_gate.html

GOSIF: https://data.globalecology.unh.edu/data/GOSIF_v2